\documentclass[aps,amsmath,asssymb,superscriptaddress,groupedaddress]{revtex4}  % for review and submission
\usepackage{graphicx}  % needed for figures
\usepackage{dcolumn}   % needed for some tables
\usepackage{bm}        % for math
\usepackage{amssymb}   % for math
 \usepackage{slashed}
\usepackage{epsf}
\usepackage{hyperref}
\usepackage{enumitem}
\usepackage{lipsum}

\newcommand{\bea}{\begin{eqnarray}}
\newcommand{\eea}{\end{eqnarray}}
\newcommand{\la}{\label}
\newcommand{\be}{\begin{equation}}
\newcommand{\ee}{\end{equation}}
\newcommand{\tr}{\,\mbox{tr}\,}

\begin{document}

\title{ Double Spin Asymmetries through  QCD Instantons}

\author{Yachao Qian}
\address{Department of Physics and Astronomy,
Stony Brook University,  Stony Brook, NY 11794-3800.}

\author{Ismail Zahed}
\address{Department of Physics and Astronomy,
Stony Brook University,  Stony Brook, NY 11794-3800.}

%\date{\today}

\begin{abstract}
We revisit the large instanton contribution to the gluon Pauli form factor of the constituent quark
noted by Kochelev. We check that it contributes sizably to the single spin asymmetry in polarized
$p_\uparrow p \rightarrow \pi X$. We use it to predict a large double
spin asymmetry in doubly polarized $p_\uparrow p_\uparrow\rightarrow \pi \pi X$. 
\end{abstract}

\maketitle

%%%%%%%%%%%%%%%%%%%%%%%%%%%%%%
%%%%%%%%%%%%%%%%%%%%%%%%%%%%%%
\section{\label{sec:introduction}introduction}
%%%%%%%%%%%%%%%%%%%%%%%%%%%%%%
%%%%%%%%%%%%%%%%%%%%%%%%%%%%%%
The QCD vacuum is dominated by large instanton and anti-instanton fluctuations in the 
infrared, that are largely responsible for the spontaneous breaking of chiral symmetry
and the anomalously large $\eta^\prime$ mass~\cite{Schafer:1996wv, nowak1996chiral}. QCD instantons
may contribute substantially to small angle hadron-hadron scattering~\cite{shuryak2000,Nowak2001,shuryak2004,kharzeev2000,Dorokhov:2004rj} 
and possibly gluon saturation at HERA~\cite{ringwald2001,schrempp2003}, as
evidenced by recent  lattice investigations~\cite{Giordano:2009vs,Giordano:2011zv}.

A number of semi-inclusive DIS experiments carried by the CLAS and HERMES collaborations 
~\cite{hermes2000,hermes2005,hermes2009,clas2010}, and more recently with polarized
protons on protons by the STAR and PHENIX collaborations~\cite{star2008,Eyser:2006bc, fnal1991},
have revealed large spin asymmetries  in polarized lepton-hadron and hadron-hadron
collisions at collider energies.  These effects  are triggered by T-odd contributions in the
scattering amplitude.

Perturbative QCD does not support the T-odd contributions,  which are usually
parametrized in the initial state (Sivers effect)~\cite{sivers1990,sivers1991} or 
the final state (Collins effect)~\cite{collins1993,collins1994}. Non-perturbative
QCD with instantons allow for large spin asymmetries as discussed by Kochelev
and others~\cite{Kochelev:1999nd,dorokhov2009,ostrovsky2005,Qian:2011ya}.  In~\cite{Kochelev:1999nd} a particularly large Pauli form factor was noted, with 
an important contribution to the Single Spin Asymmetry (SSA) in polarized proton on proton scattering.
In this note we would like to point out that it contributes significantly to 
doubly polarized proton on proton scattering.

The organization of the paper is as follows: In section \ref{sec:antvertex} we review the emergence of a large Pauli
form factor on a constituent quark in the QCD vacuum.  In section~\ref{sec:ssa} we assess  its effect on the transverse 
SSA  in $p_\uparrow p\rightarrow \pi^{\pm, 0} X$ following a recent argument in~\cite{Kochelev:2013zoa}. 
The effect is comparable in magnitude to the one discussed in~\cite{ostrovsky2005,Qian:2011ya}. 
In section~\ref{sec:DSA} we show that it gives a subtantial contribution to the Double Spin Asymmetry (DSA) 
in semi-inclusive  $p_\uparrow p_\uparrow\rightarrow \pi \pi X$. Our conclusions follow in section~\ref{sec:summary}.

 %%%%%%%%%%%%%%%%%%%%%%%%%%%%%%
\section{  Effective Pauli Form Factor   }
\la{sec:antvertex}
%%%%%%%%%%%%%%%%%%%%%%%%%%%%%%

\begin{figure}
\includegraphics[height=40mm]{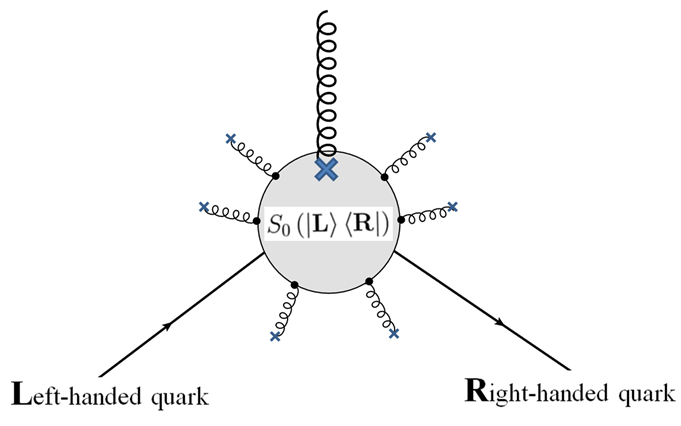}
\caption{ Effective Quark-Gluon vertex in the instanton vacuum.}
\label{backquantumvertex}
\end{figure}

The QCD vacuum is a random ensemble of  instantons and anti-instantons interacting via the exchange of
perturbative gluons and quasi-zero modes of light quarks and anti-quarks. In the dilute instanton approximation,
a typical effective vertex with quarks and gluons attached to an instanton is shown in Fig.~\ref{backquantumvertex}.
The corresponding effective vertex is given by~\cite{ 'tHooft:1976fv, Vainshtein:1981wh, Kochelev:1996pv},

\bea\la{antvertexeq1}
\mathcal{L} = \int \prod_q \left[ m_q \rho - 2 \pi^2 \rho^3 \bar{q}_R \left( 1 + \frac{i}{4} \tau^a \bar{\eta}_{\mu\nu}^a \sigma_{\mu\nu} \right) q_L \right] \exp\left( - \frac{2 \pi^2}{g_s} \rho^2 \bar{\eta}^b_{\gamma \delta} G^b_{\gamma \delta} F_g(\rho Q) \right) d_0 (\rho) \frac{d \rho}{\rho^5} d \bar{\sigma} + \left( L \leftrightarrow R \right)
\eea
where $d \bar{\sigma}$ is the integration over the instanton orientation in color space and  $\sigma_{\mu\nu} = [\gamma_\mu ,\gamma_\nu]/2$.  The incoming and outgoing quarks have small momenta $p$ ($\rho p << 1$) and $Q$ is the momentum
transferred by the inserted gluon with a form-factor

\be
F_g (x) \equiv \frac{4}{x^2} - 2 K_2 (x) \xrightarrow{x \rightarrow 0} 1
\ee
By expanding Eq.~\ref{antvertexeq1} to leading order in the inserted gluon field 
of $G^b_{\gamma \delta}$ and integrating  over the color indices, we obtain

\be
    \frac{i}{g_s}  F_g(\rho Q)   \int   \pi^4 \rho^4 \frac{ \bar{q}_R   t^a   \sigma_{\mu\nu}   q_L }{m_q^*}   G^a_{\mu\nu} \times  \left( \prod_q \left( \rho m_q^* \right) d_0 (\rho) \frac{d \rho}{\rho^5} \right)  =   \frac{i}{g_s}    F_g(\rho Q)   \int   d \rho~\pi^4 \rho^4   n(\rho) \frac{ \bar{q}_R   t^a   \sigma_{\mu\nu}   q_L }{m_q^*}   G^a_{\mu\nu}   
\ee
where $n(\rho)$ is the effective instanton density and $m^*_q$ is the effective quark mass. 
In the dilute instanton approximation~\cite{Shuryak:1981fza}

\be
n(\rho) = n_I \delta (\rho - \rho_c)
\ee
where $\rho_c$ is the average size of the instanton.  Hence the induced instanton effective quark-gluon vertex

\be\la{antvertexeq2}
 \frac{i}{g_s}   F_g(\rho Q)   \pi^4   (n_I \rho^4_c)   \frac{ \bar{q}_R   t^a   \sigma_{\mu\nu}   q_L }{m_q^*}   G^a_{\mu\nu}   
\ee
as illustrated in Fig.~\ref{backquantumvertex}.  In momentum space, the effective vertex is $M_\mu^a$ and reads
\be\la{effectivevertex}
M_\mu^a = \gamma_\mu t^a -   \frac{2}{g_s^2}   F_g(\rho Q)   \pi^4   (n_I \rho^4_c)   \frac{     t^a   \sigma_{\mu\nu}   }{m_q^*}   q^\nu 
\ee
after analytical continuation to Minkowski Space. Eq.~\ref{antvertexeq2} yields an anomalously large  Quark Chromomagnetic Moment~\cite{Kochelev:1996pv}

\be
\mu_a = - \frac{2 n_I \pi^4 \rho_c^4}{g_s^2}
\ee

%%%%%%%%%%%%%%%%%%%%%%%%%%%%%%
\section{Single Spin Asymmetries}
\la{sec:ssa}
%%%%%%%%%%%%%%%%%%%%%%%%%%%%%%

%%%%%%%%%%%%%%%%%%%%%%%%%%%%%%
\subsection{SSA: Estimate}
\la{sec:comparison}
%%%%%%%%%%%%%%%%%%%%%%%%%%%%%%

\begin{figure}[!htb]
\minipage{0.45\textwidth}
\includegraphics[height=35mm]{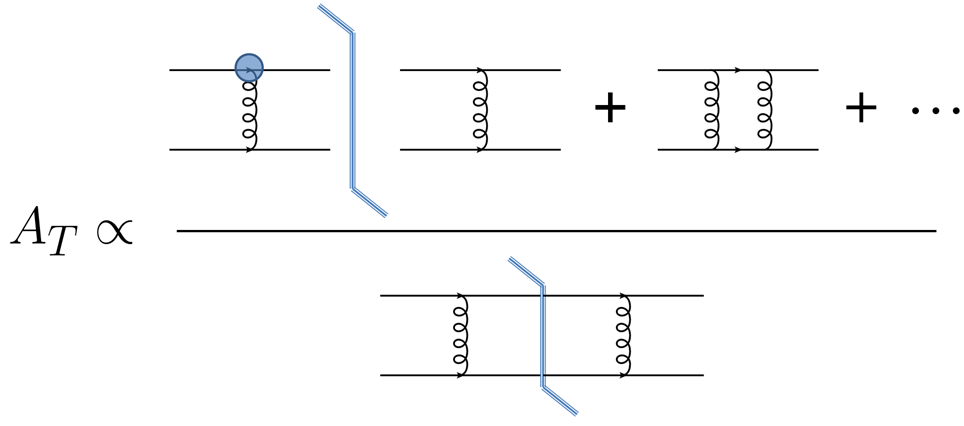}
\caption{Schematically diagrammatic contributions to the SSA through the Pauli Form factor~\cite{Kochelev:2013zoa} \la{ssaexpansion} }
\endminipage\hfill
\minipage{0.45\textwidth}
\includegraphics[height=35mm]{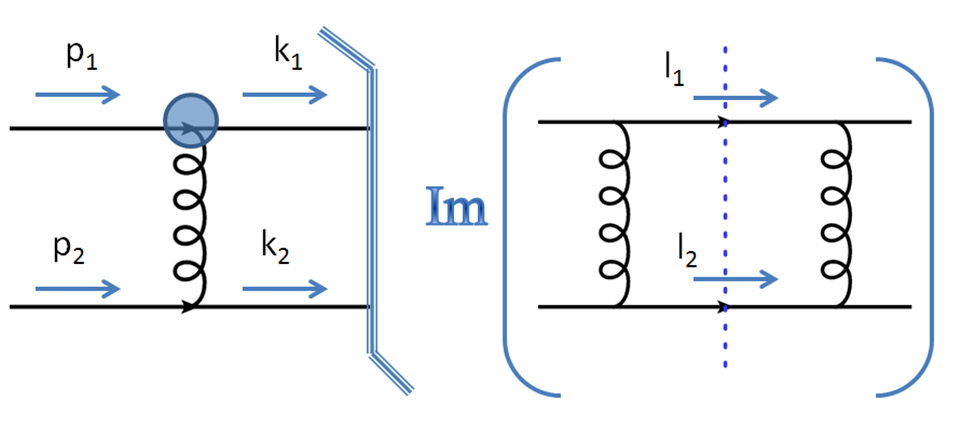}
\caption{ Leading diagrammatic contribution to the SSA through the Pauli form factor.  \la{ssaimaginary} }
\endminipage
\end{figure}

To calibrate the effects of Eq.~\ref{effectivevertex} on the double spin asymmetries, we revisit its contribution to
the SSA on semi-inclusive and polarized $p_\uparrow p\rightarrow \pi^{\pm, 0} X$ experiments.  This effect was
recently discussed in~\cite{Kochelev:2013zoa}, so we will be brief.
In going through an instanton, the chirality of the light quark can be flipped. Using the Pauli form factor 
discussed in Sec-\ref{sec:antvertex}, the SSA follows from the diagrams of Fig.~\ref{ssaexpansion}. 
As noted in \cite{Kochelev:2013zoa}, the leading diagram contributing to the SSA is displayed in Fig.~\ref{ssaimaginary}. 
Note that Fig.~\ref{ssaimaginary} is of the same order in $g_s$ as the zeroth order diagram in Fig.~\ref{ssaexpansion},  since
the chirality-flip effective vertex (Eq.~\ref{effectivevertex}) is semi-classical and of order $1/g_s^2$. 
The zeroth order differential cross section reads

\be\la{zeroorder}
d^{(0)} \sigma  \sim     \frac{64 g_s^4}{|p_1 - k_1|^4} [(k_1 \cdot p_2) (k_2 \cdot p_1) + (k_1 \cdot k_2) (p_1 \cdot p_2)]
\ee
The  first order differential cross section for the chirality flip reads~\cite{Potter:1997za}
\be\la{firstordereq1}
d^{(1)} \sigma \sim  i \frac{g_s^6}{(k_1 - p_1)^2 } \frac{1   }{ 16\pi}  \frac{(4 \pi)^\epsilon}{\Gamma(1 - \epsilon)} \frac{\mu^{2 \epsilon}}{s^\epsilon} \int_0^1 d y~[y (1-y)]^{- \epsilon} \int_0^{2 \pi}\frac{ d \phi_l}{2 \pi}  \frac{1}{(l_1 - k_1)^2 } \frac{1}{(p_1 - l_1)^2 }\mathcal{G} (\Omega)
\ee
where  $y = (1+\cos \theta_l)/2$, $\pm \theta_l$ is the longitudinal angle of $l_{1/2}$ and
\be
\mathcal{G} (\Omega) \equiv  \tr[ (M_\mu^a)^{(1)}  \slashed{p}_1  \gamma_5 \slashed{s} \gamma_\nu t^b \slashed{l}_1 \gamma_\rho  t^c \slashed{k}_1  ] \tr [   \gamma^\mu t_a \slashed{p}_2 \gamma^\nu t_b \slashed{l}_2  \gamma^\rho t_c \slashed{k}_2] 
\ee
From Sec-\ref{sec:antvertex}
\be
(M_\mu^a)^{(1)} = -   \frac{   t^a   }{g_s^2}   \frac{    F_g(\rho Q)   \pi^4   (n_I \rho^4_c)     }{m_q^*}  [\gamma_\mu (\slashed{k}_1 - \slashed{p}_1) + (\slashed{p}_1 - \slashed{k}_1)\gamma_\mu ]
\ee

To simplify the analysis and compare to existing semi-inclusive data, we use the  kinematics 
\bea\la{kinematicssimple}
 p_{1/2} &=&  \frac{\sqrt{\tilde{s}}}{2} (1, 0, 0, \pm 1) \nonumber\\
k_{1/2} &=&   \frac{\sqrt{\tilde{s}}}{2}  (1,  \pm \sin \theta  \sin \phi  ,  \pm \sin \theta   \cos \phi  , \pm \cos \theta  ) \nonumber\\
s  &=& (0,0, s^\perp , 0) 
\eea
where   $\sqrt{\tilde{s}}$ is the total energy of the colliding "partons". It is  simple to show that $d^{(1)} \sigma \sim \vec{k}_1 \cdot ( \vec{p}_1 \times \vec{s}) \sim \sqrt{\tilde{s}} s^\perp k^\perp_1 \sin \phi$, which results in SSA. For simplicity, we calculate the first differential cross section $d^{(1)} \sigma$ with $\phi = \pi/2$, where the transverse momentum of the outgoing particle lines along the $x$ axis.  Straightforward algebra yields
\bea
d^{(1)} \sigma &\sim&  s^\perp k^\perp_1 \frac{2 g_s^4 }{3 \pi} \frac{\Gamma(- \epsilon)}{\Gamma(2 - 2\epsilon) \Gamma(1 - \epsilon)}   \csc^2 (\theta)   (4 \pi)^\epsilon  \frac{\mu^{2 \epsilon}}{s^\epsilon}
  \frac{    F_g(\rho Q)   \pi^4   (n_I \rho^4_c)     }{m_q^*}   \nonumber\\
&& \times [25 \epsilon - 12 + \cos \theta (\epsilon (9 + 2  \epsilon) - 4)   {}_2F_1 (1, 1- \epsilon, 1 - 2 \epsilon, \sec^2 \frac{\theta}{2} )   + \epsilon (1 - \cos \theta) {}_2F_1 (2, 1- \epsilon, 1 - 2 \epsilon, \sec^2 \frac{\theta}{2} )    ]\nonumber\\
\eea
where ${}_2F(a,b,c; y)$ is a  hypergeometric function. We note that $ |{}_2F_1 (1,1,1;y) |$ is much larger than $| {}_2F_1^{(0,1,0,0)}(1,1,1;y)|$ and $|{}_2F_1^{(0,0,1,0)}(1,1,1;y)|$  for $y \sim1$. Therefore
\be \la{typetworesult}
 d^{(1)} \sigma     \sim     s^\perp k^\perp_1 \frac{2 g_s^4 }{3 \pi}   \frac{    F_g(\rho Q)   \pi^4   (n_I \rho^4_c)     }{m_q^*}   \csc^4 ( \frac{\theta}{2})  (3 + \cos \theta ) \left( - \frac{1}{\epsilon} + 2 \gamma_{\rm E} + \ln (\frac{\tilde{s}}{4 \pi \mu^2})  \right) 
\ee 
The divergence in (\ref{typetworesult})  stems from the exchange of soft gluons in the box diagram.
In~\cite{Kochelev:2013zoa} it was regulated using a constituent gluon mass $m_g$. For $\theta_l \sim 0$,   $\vec{l}_1$ is parallel to   $\vec{p}_1$, and this collinear divergence could be regulated by restricting $- (l_1 - p_1)^2 > m^2_g$ or equivalently setting $y_{\rm max} \sim 1-  c\, {m_g^2}/{\tilde{s}}$ with
$c$ an arbitrary constant of order 1. This regularization amounts to the substitution

\be\la{reducedinte}
\int_0^1 dy \longrightarrow \left( \int_0^{\frac{1 + \cos \theta}{2} - c \frac{m_g^2}{\tilde{s}}}  +\int_{\frac{1 + \cos \theta}{2}+ c \frac{m_g^2}{\tilde{s}}}^{1- c \frac{m_g^2}{\tilde{s}}} \right)   dy
\ee
in Eq.~\ref{firstordereq1},
where we have also regulated the collinear divergence when  $\vec{l}_1$ is parallel to $\vec{k}_1$. Thus

\be
 \left( - \frac{1}{\epsilon} + 2 \gamma_{\rm E} + \ln (\frac{\tilde{s}}{4 \pi \mu^2})  \right)  \longrightarrow  \ln \left( c \frac{\tilde{s}}{m_g^2} \right)    + \ln\left( \frac{1 - \cos \theta}{1 + \cos \theta} \right)
\ee
The regulated SSA is now given by

\be
A_{T}^{\sin \phi} \approx \frac{  d^{(1)} \sigma  }{    d^{(0)} \sigma  } = s^\perp k^\perp_1 \frac{    F_g\left( \rho Q \right)  \pi^3   (n_I \rho^4_c)     }{m_q^*} \frac{      (3 + \cos \theta )     }{6  (5 + 2 \cos \theta + \cos^2 \theta)} \left[  \ln \left(   c \frac{\tilde{s}}{m_g^2} \right)    + \ln\left( \frac{1 - \cos \theta}{1 + \cos \theta} \right)    \right]
\ee
where the zeroth order cross section in Eq.~\ref{zeroorder} is used for normalization.

%%%%%%%%%%%%%%%%%%%%%%%%%%%%%%
\subsection{SSA: Experiment}
\la{sec:comparison}
%%%%%%%%%%%%%%%%%%%%%%%%%%%%%%

To compare with the semi-inclusive data on $p_\uparrow p\rightarrow \pi X$,  we set $s^\perp u(x,Q^2) = \Delta_s u (x,Q^2)$ and $s^\perp d(x,Q^2) = \Delta_s d (x,Q^2)$, with $\Delta_s u (x,Q^2)$ and $\Delta_s d (x,Q^2)$ as the spin polarized distribution functions of the valence up-quarks and valence down-quarks in the proton respectively.    For forward $\pi^+$, $\pi^-$ and $\pi^0$ productions, the SSAs are

\be\la{piplus}
 A_{T}^{\sin \phi} (\pi^+) =   \left( n_I \frac{  \rho^4}{m^*_q}  \right) k^\perp  \frac{\Delta_s u(x_1  ,Q^2)}{u(x_1  , Q^2)}  \pi^3  F_g \left( \rho Q \right) \frac{   (3 +\cos \theta)   }{6  (5 + 2\cos \theta  + \cos^2 \theta)}\left[  \ln \left(   c \frac{\tilde{s}}{m_g^2} \right)    + \ln\left( \frac{1 - \cos \theta}{1 + \cos \theta} \right)    \right]
\ee

\be\la{piminus}
 A_{T}^{\sin \phi} (\pi^-) = \left( n_I \frac{  \rho^4}{m^*_q}  \right)  k^\perp  \frac{\Delta_s d(x_1 ,Q^2)}{d(x_1  , Q^2)}  \pi^3  F_g \left( \rho Q \right) \frac{   (3 +\cos \theta)   }{6  (5 + 2\cos \theta  + \cos^2 \theta)}\left[  \ln \left(   c \frac{\tilde{s}}{m_g^2} \right)    + \ln\left( \frac{1 - \cos \theta}{1 + \cos \theta} \right)    \right]
\ee

\be\la{pizero}
 A_{T}^{\sin \phi} (\pi^0) = \left( n_I \frac{  \rho^4}{m^*_q}  \right)  k^\perp   \frac{\Delta_s u(x_1  ,Q^2) + \Delta_s d(x_1 ,Q^2)}{u(x_1  , Q^2) + d(x_1  , Q^2)}  \pi^3  F_g \left( \rho Q \right) \frac{   (3 +\cos \theta)   }{6  (5 + 2\cos \theta  + \cos^2 \theta)}\left[  \ln \left(   c \frac{\tilde{s}}{m_g^2} \right)    + \ln\left( \frac{1 - \cos \theta}{1 + \cos \theta} \right)    \right]
\ee 
According to \cite{Hirai:2006sr, Adams:1997dp}
\bea
  \frac{\Delta_s u(x  ,Q^2)}{u(x  , Q^2)} &=& 0.959 - 0.588 (1 - x^{1.048}) \nonumber\\
   \frac{\Delta_s d(x  ,Q^2)}{d(x  , Q^2)} &=& - 0.773 + 0.478 (1- x^{1.243})\nonumber\\
 \frac{u(x  , Q^2)}{d(x  , Q^2)} &=& 0.624 (1-x)
\eea

\begin{figure}[!htb]
\minipage{0.48\textwidth}
\includegraphics[height=40mm]{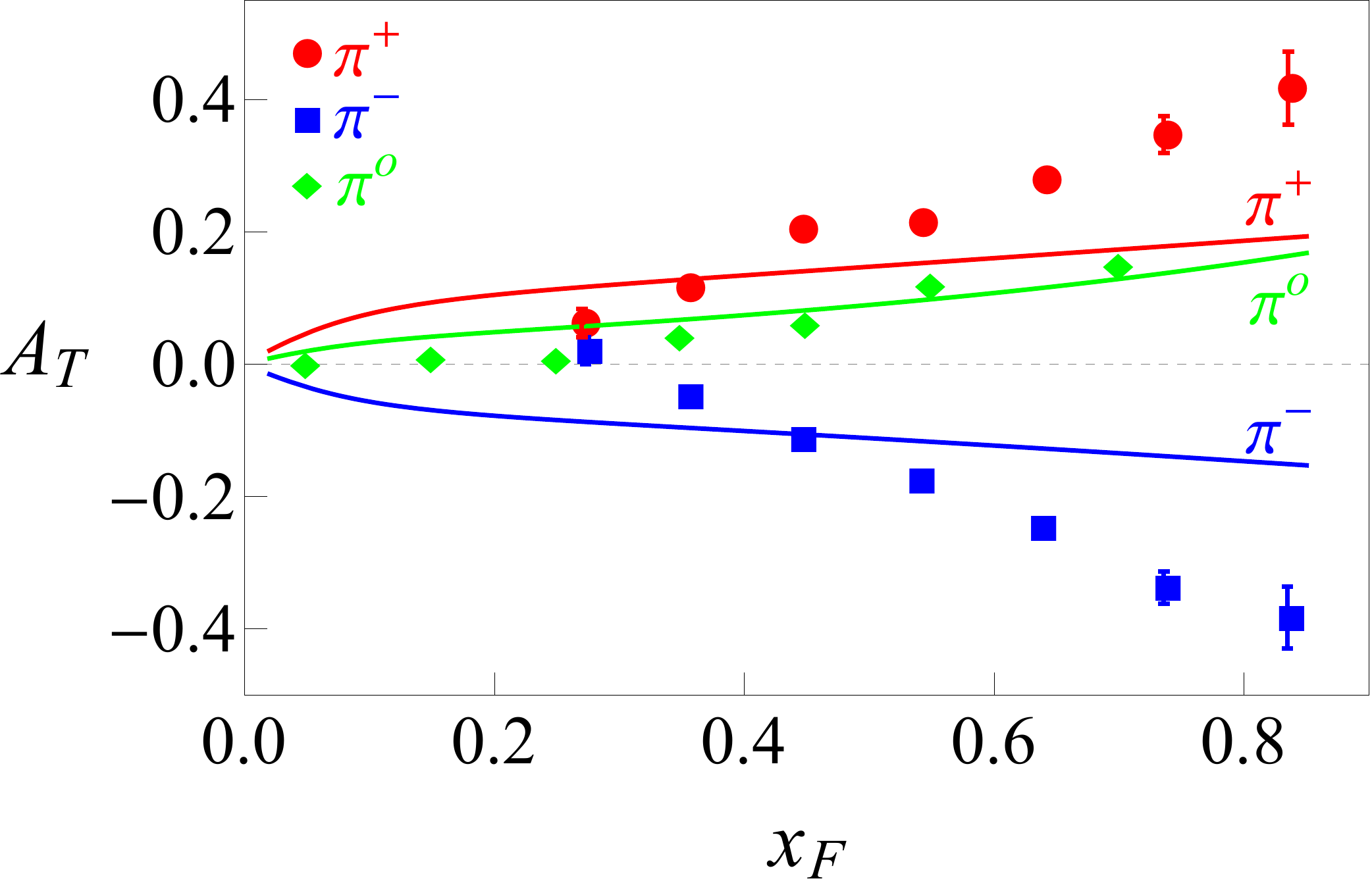}
\endminipage\hfill
\minipage{0.48\textwidth}
\includegraphics[height=40mm]{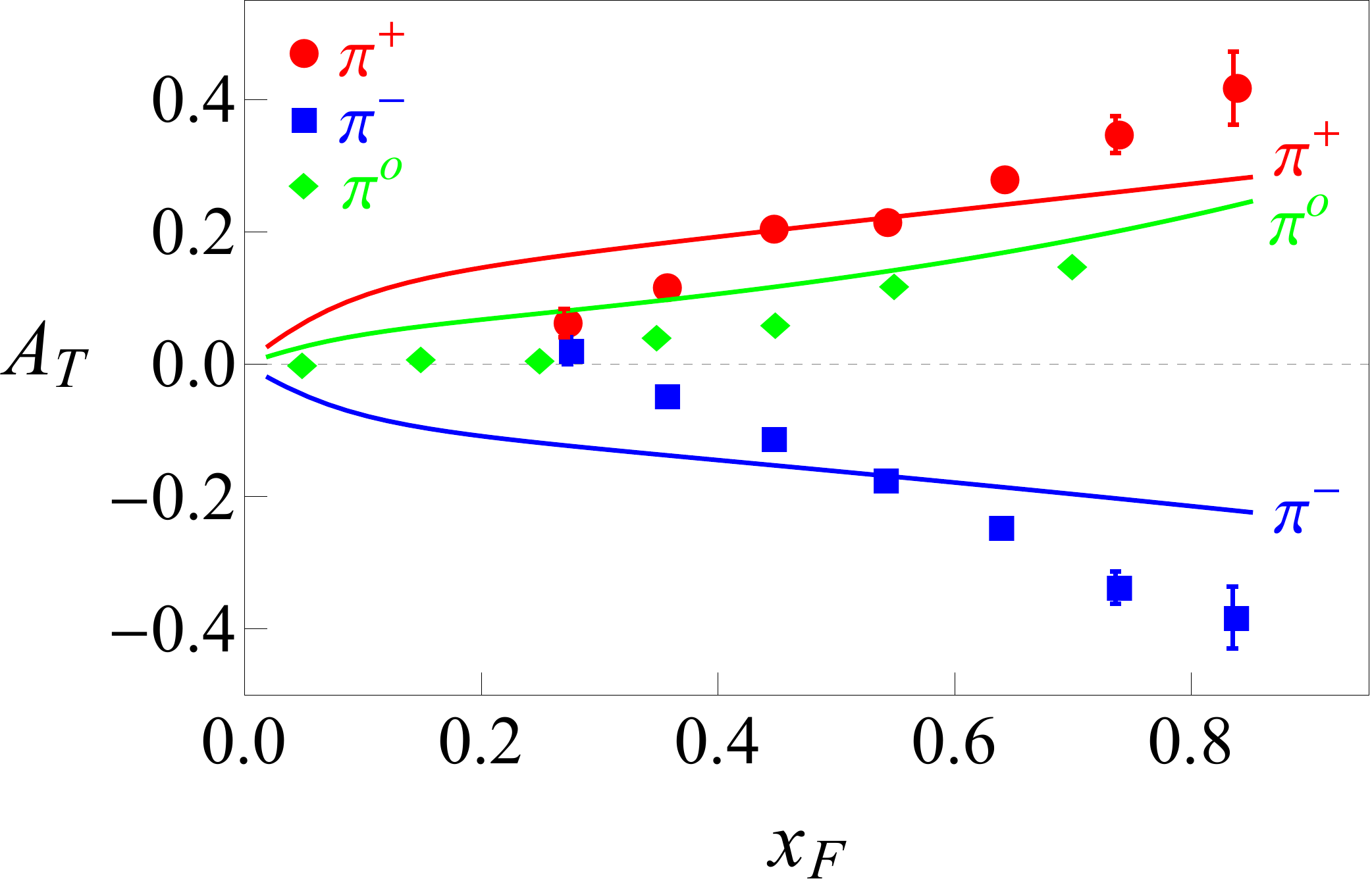}
\endminipage
  \caption{ $x_F$ dependent SSA in $p_\uparrow p\rightarrow \pi X$ collisions at $\sqrt{s}=19.4 {\rm GeV}$~\cite{Skeens:1991my}.  The solid lines are the analytical results in Eq.~\ref{piplus}- Eq.~\ref{pizero} with $c = 2$ (left) and Eq.~\ref{NUM} (right). }\label{ssafigure}
\end{figure}

These results can be compared to the experimental measurements in~\cite{Skeens:1991my}. For simplificty,  we assume the same fraction for each proton $\left<x_1\right> = \left<x_2\right>  = \left<x \right> $,  and $\left<k^\perp \right> \approx \left<K_\perp\right>$ is the transverse momentum of the outgoing pion. We then have $ \sqrt{s} \left<x\right> \left<\sin \theta \right>   = 2 \left<K_\perp\right>$ and $\left<x \right> \left<\cos\theta \right> = \left<x_F\right>$.   For large $\sqrt{s}$, we also have $\left<Q\right> \approx \left<K_\perp \right> \sqrt{\left<x \right>/ \left<x_F \right>}$. We set $c = 2$ and $\left< K_\perp \right> = 2 {\rm GeV}$ for the outgoing pions.  $n_I \approx 1/{\rm fm}^4$ is the effective instanton density, $\rho \approx 1/3 {\rm fm}$ the typical instanton size and $m_q^* \approx 300 {\rm MeV}$ the constitutive quark mass in the instanton vacuum.    $m_g \approx 420 {\rm MeV}$ is the effective gluon mass in the instanton vacuum\cite{Hutter:1993sc}. In Fig.~\ref{ssafigure} (left) we  display the
results (\ref{piplus}-\ref{pizero}) as a function of the parton fraction $x_F$ for both the charged
and uncharged pions at $\sqrt{s}=19.4\,{\rm GeV}$~\cite{Skeens:1991my}. 

Instead of isolating the collinear divergence, we can also numerically compute Eq.~\ref{firstordereq1} with a massive gluon propagator
as was discussed also in~\cite{Kochelev:2013zoa}

\be
\label{NUM}
d^{(1)} \sigma \sim  i \frac{g_s^6}{(k_1 - p_1)^2 - m_g^2} \frac{1   }{ 16\pi}  \int_0^1 d y   \int_0^{2 \pi}\frac{ d \phi_l}{2 \pi}  \frac{1}{(l_1 - k_1)^2- m_g^2 } \frac{1}{(p_1 - l_1)^2 - m_g^2}\mathcal{G} (\Omega)
\ee
The numerical results are displayed in Fig.~\ref{ssafigure} (right). Both regularizations lead about similar results. 
In sum, the anomalous Pauli form factor
can reproduce the correct magnitude of the observed SSA in 
polarized $p_\uparrow p\rightarrow \pi X$ for reasonable vacuum parameters. The contribution
is comparable in magnitude to the one recently discussed in~\cite{Qian:2011ya} (see Fig.~11 in p.7) using instead
the standard Dirac form factor but changes in the instanton distribution due to the polarized
proton.

%%%%%%%%%%%%%%%%%%%%%%%%%%%%%%
%%%%%%%%%%%%%%%%%%%%%%%%%%%%%%
\section{Double Spin Asymmetries in Dijet Productions}
\la{sec:DSA}
%%%%%%%%%%%%%%%%%%%%%%%%%%%%%%
%%%%%%%%%%%%%%%%%%%%%%%%%%%%%%

 \subsection{DSA: Estimate}
 
\begin{figure}
\includegraphics[height=40mm]{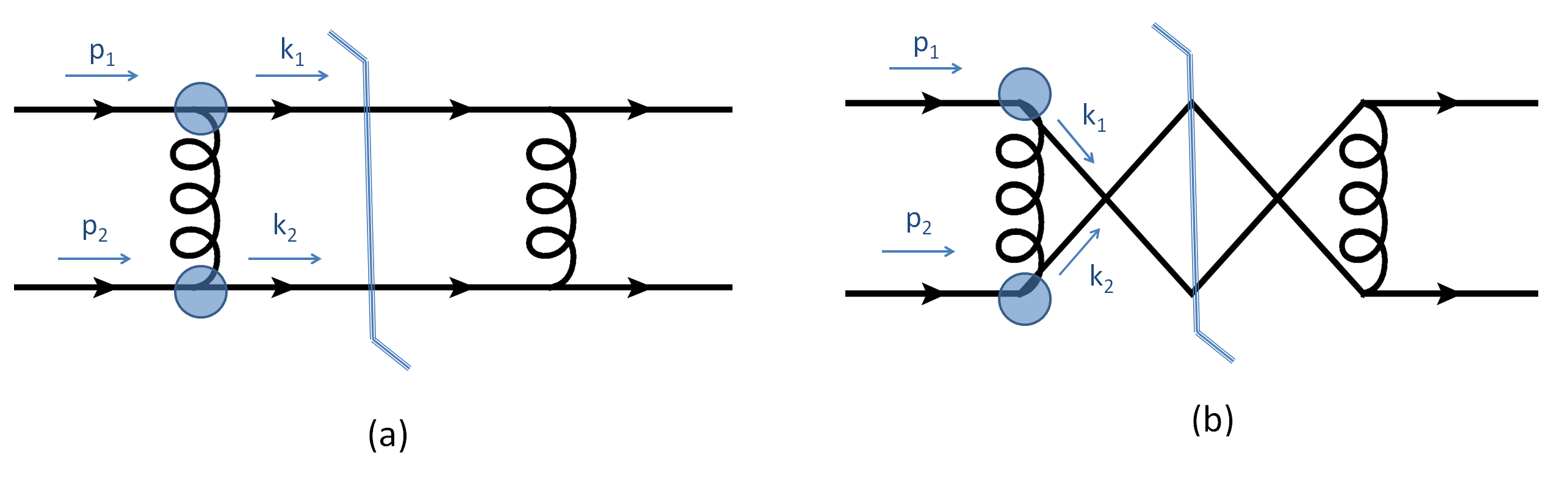}
\caption{ The valence quark in polarized proton $p_1$ exchange one gluon  with the valence quark in the polarized proton $p_2$.}
\label{dspphardgluon}
\end{figure}

The same Pauli form factor and vacuum parameters can be used to assess the role of the
QCD instantons on doubly polarized and semi-inclusive $p_\uparrow p_\uparrow\rightarrow \pi\pi X$
processes. The Double Spin Asymmetry (DSA) is defined as 

\be\la{doublespinasymmetry}
A_{\rm DS}= \frac{\sigma^{\uparrow \uparrow + \downarrow \downarrow} - \sigma^{\downarrow \uparrow + \uparrow \downarrow}}{\sigma^{\uparrow \uparrow + \downarrow \downarrow} + \sigma^{\downarrow \uparrow + \uparrow \downarrow}}
\ee
with the proton beam  polarized along the transverse direction. 
The valence quark from the polarized proton $P_1$ exchanges one gluon with the valence quark from the polarized proton $P_2$  as shown in~Fig.~\ref{dspphardgluon}.  At large $\sqrt{s}$,  Fig.~\ref{dspphardgluon}-(a) is dominant in forward pion production and  Fig.~\ref{dspphardgluon}-(b) is dominant in backward pion production. For Fig.~\ref{dspphardgluon}-(a), the differential cross section reads

\be 
d \sigma  \sim    \frac{g_s^4}{|p_1 - k_1|^4} \sum_{\rm color}  \tr[ M_\mu^a  \slashed{p}_1 (1 + \gamma_5 \slashed{s}_1)   \gamma_0 ( M_{\nu}^b  )^\dagger \gamma_0    \slashed{k}_1 ]    \tr[ M_\mu^a  \slashed{p}_2 (1 + \gamma_5 \slashed{s}_2)   \gamma_0 ( M_{\nu}^b  )^\dagger \gamma_0   \slashed{k}_2]   
\ee 
Using the anomalous Pauli form factor (\ref{effectivevertex}), the contribution to the DSA  follows from simple algebra

\bea
\label{d2}
d^{(2)} \sigma  \sim  &&   \frac{256  }{|p_1 - k_1|^4} \left(  \frac{F_g(\rho Q)   \pi^4  n_I \rho^4_c}{m_q^* } \right)^2  [  (k_1\cdot s_1) (k_1\cdot s_2) (k_2\cdot p_1) (k_2\cdot p_2)
-(k_1\cdot p_1) (k_1\cdot s_2) (k_2\cdot p_2) (k_2\cdot s_1) \nonumber\\
&&-(k_1\cdot s_1) (k_1\cdot s_2) (k_2\cdot p_2) (p_1\cdot p_2)
+(k_1\cdot k_2) (k_1\cdot p_1) (k_2\cdot p_2) (s_1\cdot s_2)
-(k_1\cdot p_1) (k_1\cdot p_2) (k_2\cdot p_2) (s_1\cdot s_2)\nonumber\\
&&-(k_1\cdot p_1) (k_2\cdot p_1) (k_2\cdot p_2 (s_1\cdot s_2) 
+(k_1\cdot p_1) (k_2\cdot p_2) (p_1\cdot p_2) (s_1\cdot s_2)
-(k_1\cdot p_2) (k_1\cdot s_1) (k_2\cdot p_1) (k_2\cdot s_2)\nonumber\\
&&+(k_1\cdot p_1) (k_1\cdot p_2) (k_2\cdot s_1) (k_2\cdot s_2)
+(k_1\cdot k_2) (k_1\cdot s_1) (k_2\cdot s_2) (p_1\cdot p_2)
-(k_1\cdot p_1) (k_2\cdot s_1) (k_2\cdot s_2) (p_1\cdot p_2)]
\eea
after using the identity

\bea
&& \tr[(\gamma_\mu \slashed{q} -  \slashed{q} \gamma_\mu) \slashed{p} \gamma_5 \slashed{s}\gamma_\nu\slashed{k}] + \tr[\gamma_\mu \slashed{p} \gamma_5 \slashed{s}(\slashed{q}\gamma_\nu - \gamma_\nu \slashed{q})\slashed{k}] \nonumber\\
&=& \tr[(\gamma_\mu \slashed{k} +  \slashed{p} \gamma_\mu) \slashed{p} \gamma_5 \slashed{s}\gamma_\nu\slashed{k}] + \tr[\gamma_\mu \slashed{p} \gamma_5 \slashed{s}(\slashed{k}\gamma_\nu + \gamma_\nu \slashed{p}) \slashed{k}] \nonumber\\
&=& 8 i \left[ p_\mu \epsilon(\nu, k,p,s) - p_\nu \epsilon (\mu,k,p,s) + (k \cdot p) \epsilon(\mu,\nu,k,s) - (k \cdot s) \epsilon(\mu,\nu,k,p)\right]
\eea
where we used $q = k-p$ and $p \cdot s = 0$ because the protons are transversely polarized.

 For a simple empirical application of (\ref{d2}) we adopt the simple kinematical set up in Eq.~\ref{kinematicssimple}. Obtain

\be
d^{(2)} \sigma  \sim    - \frac{4  }{|p_1 - k_1|^4} \left(  \frac{F_g(\rho Q)   \pi^4  n_I \rho^4_c}{m_q^* } \right)^2 \tilde{s}^3 s_1^\perp s_2^\perp (1 - \cos \theta)^2 [4 + \cos (\theta - 2 \phi) + 2 \cos (2 \phi)  + \cos (\theta + 2 \phi)]
\ee
After adding the contribution of  Fig.~\ref{dspphardgluon}-(a) and Fig.~\ref{dspphardgluon}-(b), 
and averaging over the transverse direction $\phi$, we finally  obtain
\be
\frac{d^{(2)}  \sigma}{d^{(0)}  \sigma} \sim  - 4  s_1^\perp s_2^\perp   \left(  \frac{ \pi^4  n_I \rho^4_c}{m_q^* g_s^2} \right)^2    \frac{ F_g^2 [\rho \sqrt{\frac{\tilde{s}(1 - \cos \theta)}{2}} ]  \tilde{s}   + F_g^2 [\rho \sqrt{\frac{\tilde{s}(1 + \cos \theta)}{2}} ]  \tilde{s}    }{\frac{5 + 2 \cos \theta +   \cos^2 \theta}{ (1 - \cos \theta)^2}  + \frac{5 - 2 \cos \theta +   \cos^2 \theta}{ (1 +\cos \theta)^2} }
\ee

\subsection{DSA:  Results}

\begin{figure}[!htb]
\minipage{0.33333333333 \textwidth}
\includegraphics[height=45mm]{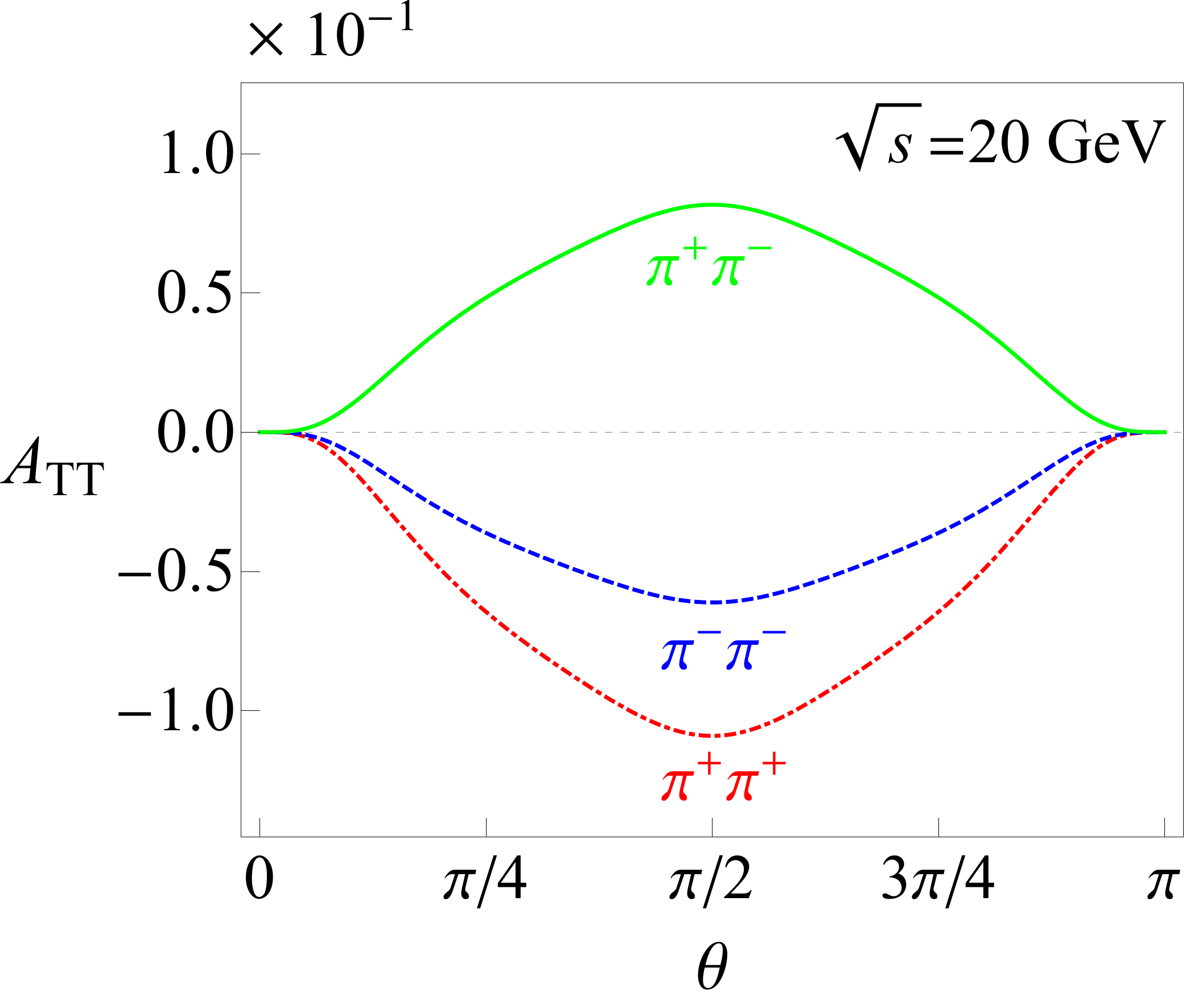}
\endminipage\hfill
\minipage{0.33333333333 \textwidth}
\includegraphics[height=45mm]{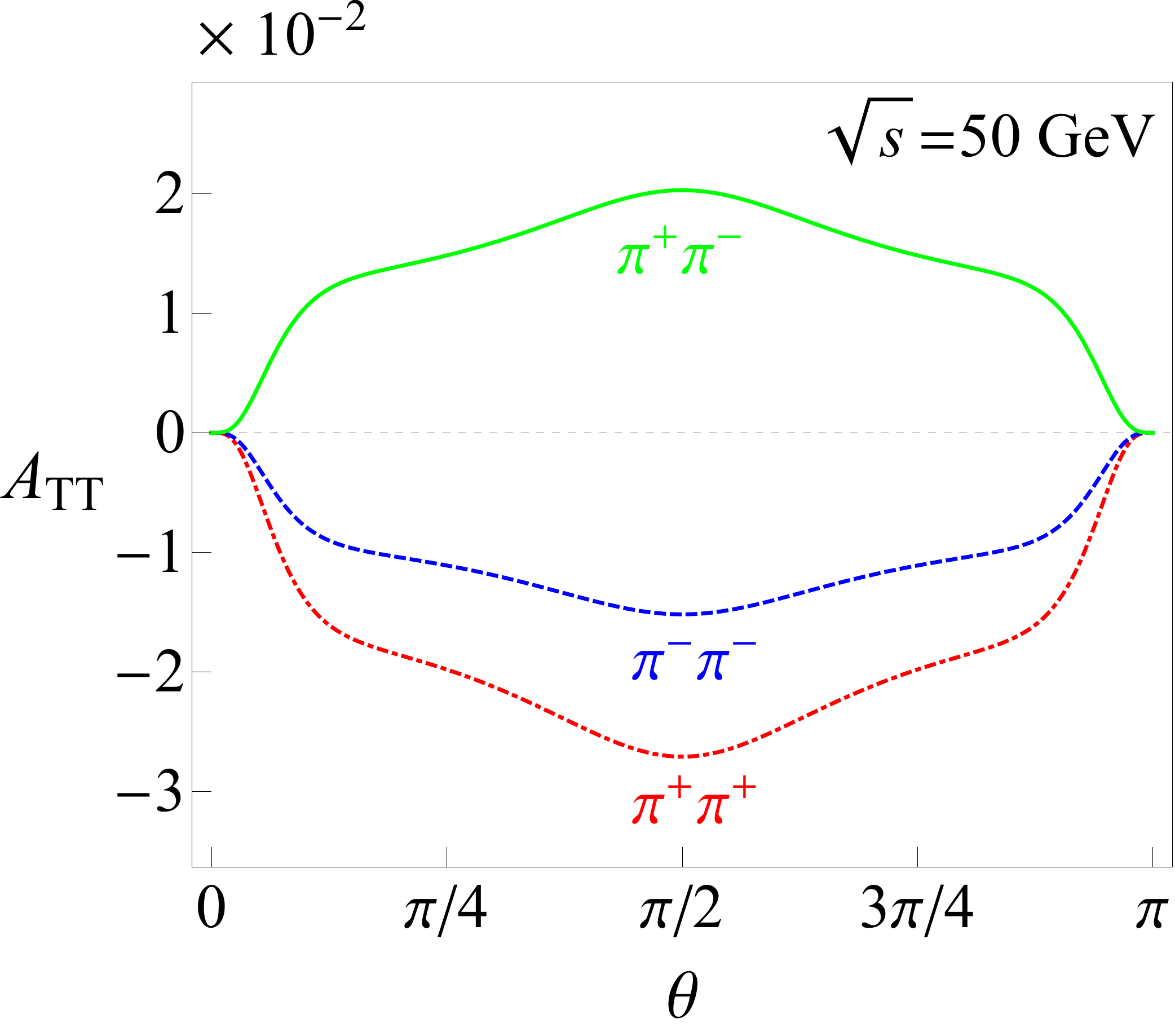}
\endminipage
\minipage{0.33333333333 \textwidth}
\includegraphics[height=45mm]{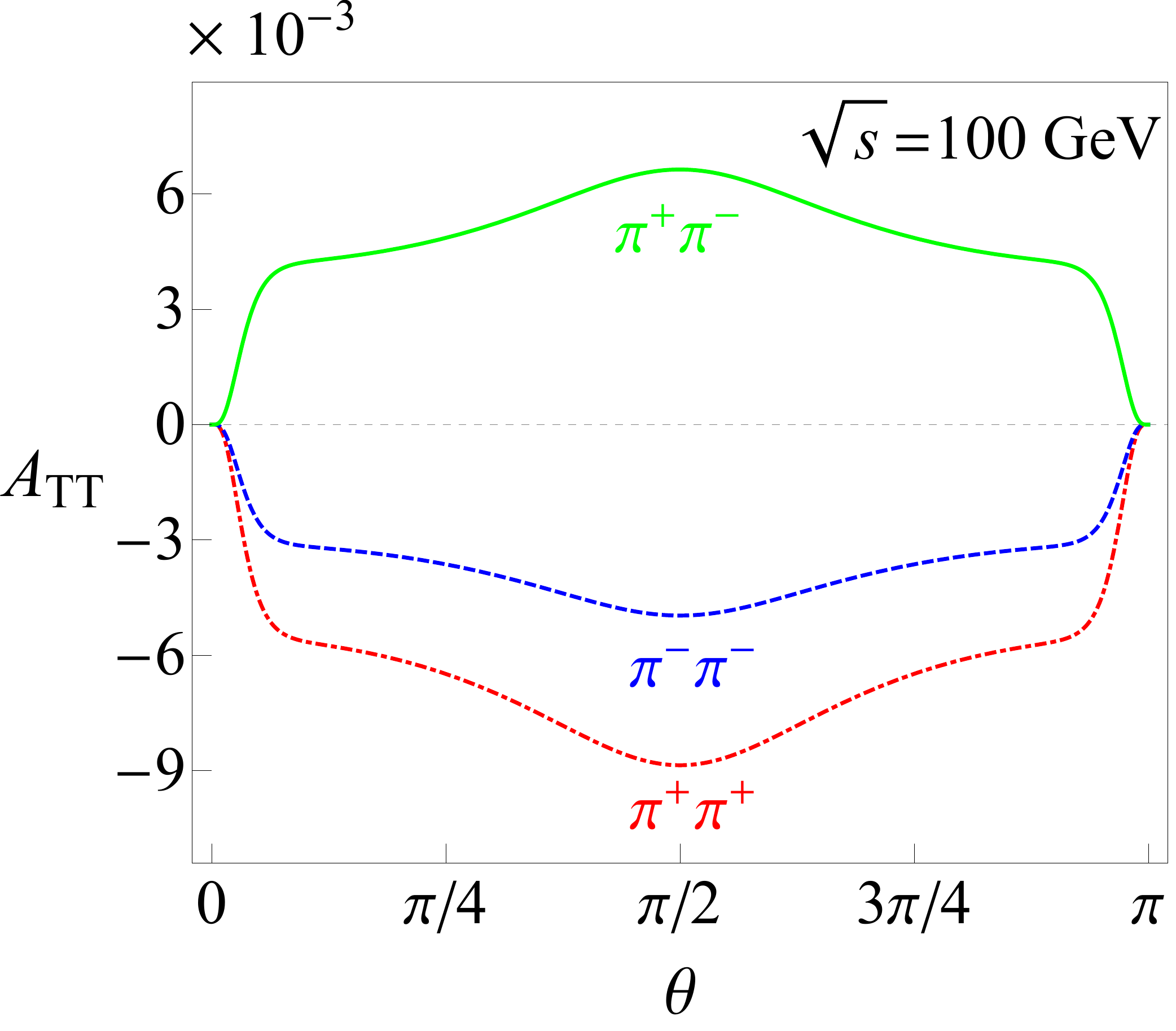}
\endminipage
  \caption{ Dotdashed  line is the Double Spin Asymmetry of $\pi^+\pi^+$ productions (Eq.~\ref{doublepiplus}). Dashed line is the Double Spin Asymmetry of $\pi^-\pi^-$ productions (Eq.~\ref{doublepiminus}). Solid line is the Double Spin Asymmetry of $\pi^+\pi^-$ productions (Eq.~\ref{doublepiplusminus}). }\label{doublegraphs}
\end{figure}

Our DSA results can now be compared to future experiments at collider energies.
Specifically, our DSA for dijet productions are

\be\la{doublepiplus}
A_{\pi^+ \pi^+} = - \frac{1}{8}  \frac{\Delta_s u(x_1  ,Q^2)}{u(x_1  , Q^2)}   \frac{\Delta_s u(x_2  ,Q^2)}{u(x_2 , Q^2)} \left(  \frac{ \pi^3  n_I \rho^4_c}{m_q^* \alpha_s }   \right)^2      \frac{ F_g^2 [\rho \sqrt{\frac{\tilde{s}(1 - \cos \theta)}{2}} ]  \tilde{s}     + F_g^2 [\rho \sqrt{\frac{\tilde{s}(1 + \cos \theta)}{2}} ]  \tilde{s}  }{ (5 + 10 \cos^2 \theta + \cos^4 \theta) \csc^4 \theta }
\ee

\be\la{doublepiminus}
A_{\pi^- \pi^-} =  - \frac{1}{8} \frac{\Delta_s d(x_1  ,Q^2)}{d(x_1  , Q^2)}   \frac{\Delta_s d(x_2  ,Q^2)}{d(x_2 , Q^2)}  \left(  \frac{ \pi^3  n_I \rho^4_c}{m_q^* \alpha_s }   \right)^2      \frac{ F_g^2 [\rho \sqrt{\frac{\tilde{s}(1 - \cos \theta)}{2}} ]  \tilde{s}     + F_g^2 [\rho \sqrt{\frac{\tilde{s}(1 + \cos \theta)}{2}} ]  \tilde{s}  }{ (5 + 10 \cos^2 \theta + \cos^4 \theta) \csc^4 \theta }
\ee

\bea\la{doublepiplusminus}
A_{\pi^+ \pi^-} =&&  -  \frac{1}{8} \frac{\Delta_s u(x_1  ,Q^2) \Delta_s d(x_2  ,Q^2)+\Delta_s d(x_1  ,Q^2) \Delta_s u(x_2  ,Q^2)}{u(x_1  , Q^2)d(x_2 , Q^2)+d(x_1  , Q^2)u(x_2 , Q^2)}   \left(  \frac{ \pi^3  n_I \rho^4_c}{m_q^* \alpha_s }   \right)^2   \nonumber\\
&& \times  \frac{ F_g^2 [\rho \sqrt{\frac{\tilde{s}(1 - \cos \theta)}{2}} ]  \tilde{s}     + F_g^2 [\rho \sqrt{\frac{\tilde{s}(1 + \cos \theta)}{2}} ]  \tilde{s}  }{ (5 + 10 \cos^2 \theta + \cos^4 \theta) \csc^4 \theta }
\eea
We will assume that each  parton carries one third of the momentum of the proton $\left<x_1\right>=\left<x_2 \right> = 1/3$,
so that  $\sqrt{\tilde{s}} = \sqrt{s}/3 $, where $\sqrt{s}$ is the total energy of the colliding protons.  We will use the 
value of $\alpha_s$ from~\cite{PhysRevD.86.010001}.   The instanton size is set to $1/3 {\rm fm}$, density $n_I = 1/{\rm fm}^4$ and $m_q^* = 300 {\rm MeV}$ as for the SSA reviewed above. Our predictions for charged dijet production
in semi-inclusive DSA are displayed in  Fig.~\ref{doublegraphs}. We note that
at $\sqrt{s} \rightarrow \infty$, the Double Spin Asymmetries in (Eq.~\ref{doublepiplus} ,  Eq.~\ref{doublepiminus}, and Eq.~\ref{doublepiplusminus})
 vanish.

\section{\label{sec:summary} Conclusions}
%%%%%%%%%%%%%%%%%%%%%%%%%%%%%%
%%%%%%%%%%%%%%%%%%%%%%%%%%%%%%

QCD instantons provide a natural mechanism for large spin asymmetries in polarized dilepton and hadron 
scattering at collider energies. A simple mechanism for these large spin asymmetries was noted by Kochelev~\cite{Kochelev:1999nd} in the form of a large Pauli form factor for a constituent quark whether through photon exchange or gluon exchange.
A simple estimate of the SSA in $p_\uparrow p\rightarrow \pi X$ production compares fairly to the 
measured charged asymmetries  in~\cite{Skeens:1991my} both in sign and magnitude, 
using the instanton vacuum parameters. We have argued that the same anomalously large Pauli form factor
yields subtantial DSA in $p_\uparrow p_\uparrow\rightarrow \pi \pi X$. We welcome future measurements of these
asymmetries at collider facilities.

%%%%%%%%%%%%%%%%%%%%%%%%%%%%%%
%%%%%%%%%%%%%%%%%%%%%%%%%%%%%%
\section{\label{sec:acknowledgements}acknowledgements}
%%%%%%%%%%%%%%%%%%%%%%%%%%%%%%
%%%%%%%%%%%%%%%%%%%%%%%%%%%%%%

This work was supported in parts by the US-DOE grant DE-FG-88ER40388.

%%%%%%%%%%%%%%%%%%%%%%%%%%%%%%
%%%%%%%%%%%%%%%%%%%%%%%%%%%%%%
%\section{Appendix}
%\la{appendix}
%%%%%%%%%%%%%%%%%%%%%%%%%%%%%%
%%%%%%%%%%%%%%%%%%%%%%%%%%%%%%

\bibliography{instantonref}

\end{document}